\documentclass[%
 reprint,
superscriptaddress,
 amsmath,amssymb,
 aps,
]{revtex4-1}

\usepackage{graphicx}
\usepackage{braket}
\usepackage{dcolumn}
\usepackage{bm}
\usepackage{hyperref}
\usepackage[T1]{fontenc}
\usepackage{lipsum}
\usepackage{float}
\usepackage{xcolor}

\begin{document}

\title{Quantum chaos in a system with high degree of symmetries}

\author{Javier de la Cruz}
\affiliation{Instituto de Ciencias Nucleares, Universidad Nacional Aut\'onoma de M\'exico, Apdo. Postal 70-543, C.P. 04510  Cd. Mx., Mexico}

\author{Sergio Lerma-Hern\'andez}
\affiliation{Facultad de F\'isica, Universidad Veracruzana, Circuito Aguirre Beltr\'an s/n, Xalapa, Veracruz 91000, Mexico}

\author{Jorge G. Hirsch} 
\affiliation{Instituto de Ciencias Nucleares, Universidad Nacional Aut\'onoma de M\'exico, Apdo. Postal 70-543, C.P. 04510  Cd. Mx., Mexico}

\date{\today}

\begin{abstract}
We study dynamical signatures of quantum chaos in one of the most relevant models in many-body quantum mechanics,
 the Bose-Hubbard model, whose high degree of symmetries  yields  a large number  of invariant subspaces and  degenerate energy levels. 
While the standard procedure to reveal signatures of  quantum chaos requires  classifying   the energy levels  according to their symmetries,  we show that  this classification is not necessary to obtain manifestation of  spectral correlations in the temporal evolution of the survival probability.   
Our  findings exhibit the survival probability as a powerful tool to detect the presence of quantum chaos, avoiding the experimental and theoretical  challenges associated with the determination  of a complete set of energy eigenstates and their symmetry classification. 
\end{abstract}

\maketitle

\section{\label{sec:level1}Introduction}

Symmetries play an important role in the description of physical systems, helping to simplify their description and temporal evolution. In classical dynamics they are useful to find the coordinates providing the simplest description, in the many-body quantum domain they allow to divide the Hilbert space in unconnected subspaces, strongly reducing the dimensionality of the problem. Systems with symmetries are far more likely to have degenerate energy levels than those without them \cite{vonNeuman(1929),Reichl(2013)}.

Quantum chaos refers to the quantum description of classically chaotic systems, their universal properties associated with Random Matrix Theory, and the presence of the former in systems with no classical analogs \cite{Reichl(2013), stockmann(2006), Gutzwiller(2013)}. The best stablished signatures of quantum chaos are its spectral statistics, which require the classification of the states by their symmetry properties \cite{Berry(1977), Bohigas(1984), Percival(1973)}. In systems with many symmetries, this process can be very demanding on the theoretical side, and very difficult to implement in experimental studies. For example in the Bose-Hubbard model (BHM), which is the system that we analyze here, the number of symmetry subspaces is at least as large as the number of sites in the lattice. 

The BHM is the simplest description of a set of spin-less bosons with interactions in a lattice. 
It was the first strongly correlated lattice model being realized with ultracold atoms and in which a quantum phase transitions was observed \cite{Greiner(2002)}. Due to the high degree of controllability and new observational tools, which enable the detection of each individual atom, the model is  employed to describe  experiments related to quantum simulation, quantum thermalization and in recent years to quantum virtual cooling \cite{Bakr(2009),Bakr(2010),Islam(2015),Kaufman(2016), Lukin(2019), Cotler(2019)}. 

The spectral properties associated with quantum chaos in the BHM were studied by Kolovsky and Buchleitner \cite{Kolovsky(2004)}.
Lubasch studied measures to explore the relation between quantum chaos and entanglement \cite{Lubasch(2009)}.  A semiclassical analysis connected with Bloch oscillations was presented in \cite{Kolovsky2016}.
Recently the out-of-time-order correlator and the Lyapunov exponent  have been studied in the model \cite{Shen(2017)} and the statistical distance between initially similar number distributions has been proposed as a reliable measure to distinguish regular from chaotic behavior in a Bose-Hubbard dimer \cite{Kidd(2019)}.

In the present work, we use the survival probability, a dynamical observable, to identify  quantum chaos in the BHM. The survival probability is the probability to find the system in its initial state at time $t$. In chaotic systems, it exhibits a correlation hole of universal character at long times, a dip below the saturation value reached at longer times which can be  described analytically employing the two-level form factor of   Random Matrix Theory  \cite{Alhassid(1992)}. This observable has been successfully employed in recent works as a quantum chaos indicator in spin systems \cite{Torres-Herrera(2017), Torres-Herrera(2017-2),Torres(2018),Torres-Herrera(2019), Schiulaz(2019)} and in atom-photon systems \cite{Lerma-Hernandez(2019)}.

 Here we show that, contrary to spectral statistics where the signatures of quantum chaos disappear when many symmetry sectors are considered together, in the survival probability the correlation hole manifests even if the whole set of symmetry sectors are included in the dynamics. 
For a single symmetry sector, signatures of quantum chaos are obtained both in the distribution of unfolded nearest neighbour energy differences  and  in the survival probability.  However,  when all the symmetry sectors are considered,  the distribution of energy differences changes from  Wigner-Dyson to  Poisson, whereas  the correlation hole in the survival probability is still  clearly seen. Analytical expressions are provided, which, besides giving  theoretical support to the persistence of the correlation hole,   describe the complete evolution of the survival probability, both in the case of one and several symmetry subspaces.     

The temporal evolution of the survival probability is naturally fuzzy due to quantum fluctuations, whose relative size compared with their average value do not diminish with the dimension of the system, they are not self-averaging at any scale \cite{Torres-Herrera(2019), Torres-Herrera(2019-2)}. For this reason ensemble averages over an energy window are required to make the correlation hole visible \cite{Lerma-Hernandez(2019)}. Recently it has been shown that such ensemble of  initial states can be prepared in a band of energy eigenstates and allowed to evolve  \cite{Yang(2020)}. This development could facilitate the experimental observation of the correlation hole.

The structure of the paper is as follows. In Sec. \ref{sec:model} we describe the BHM, its symmetries and the properties of its energy spectrum. Then we describe the dynamics of the survival probability and the correlation hole in Sec. \ref{sec:survivalprobability}, both in the case of one symmetry sector and when  the whole set of invariant subspaces is considered together. In  Sec. \ref{sec:results},  the analytical expressions derived in the  previous section are shown to describe properly  numerical results for the BHM in a chaotic regime. The way these signatures of quantum chaos disappear as we move to integrable limits is also discussed. To finish, our conclusions are given in Sec. \ref{sec:conclusions} .

\section{\label{sec:model}The Bose-Hubbard Model}

The model which we consider is the one dimensional BHM with periodic boundary conditions. This model describes the dynamics of $N$ spin-less bosons on a lattice with $L$ sites in a ring array. The Hamiltonian of this system, with $\hbar= 1$, is
 \begin{equation}
    \hat{H} = -J\sum_{l=1}^{L}\left(\hat{a}^{\dagger}_{l+1}\hat{a}_l + h.c.\right) + \frac{U}{2}\sum_{l=1}^L\hat{n}_l\left(\hat{n}_l-1\right).
    \label{eq:Hamiltonian}
\end{equation}
The operators $\hat{a}^{\dagger}_l$ and $\hat{a}_l$ are the creation and annihilation operators of one boson on the site $l$, respectively. Due to periodic boundary conditions, the index $L+1$ in the first sum should be considered as $L+1:=1$. The first term is the kinetic energy, describing the coherent tunneling  between adjacent sites with rate $J$. The second term describes the interaction energy on-site with intensity $U$, where $\hat{n}_l=\hat{a}_l^{\dagger}\hat{a}_l$ gives the number of particles on site $l$ and the total number of particles is constant, $\sum_{l=1}^L\hat{n}_l=N$.

One of the most relevant aspects of the system is that in the thermodynamic limit it presents a second order quantum phase transition, going from a Mott insulator to a superfluid phase \cite{Fisher1989,Greiner(2002),Cucchietti2007}.

\subsection{Symmetries and degenerated subspaces}

The symmetries of the BHM are described by  the dihedrical group $D_L$, which is the group of symmetries of a regular polygon with $L$ sides. One of the symmetries is related to the translational invariance of the Hamiltonian due to the periodic boundary conditions. The shift operator $\hat{S}$ is the responsible of the decomposition into different $\kappa$-subspaces and acts in a Fock state as

\begin{equation}
    \hat{S}\ket{n_1,n_2,...,n_{L-1},n_L} = \ket{n_L,n_1,n_2,...,n_{L-1}}. 
\end{equation}
The eigenvalues of $\hat{S}$ are $a_j=e^{i\kappa_j}$ with $\kappa_j=2\pi j/L$, the single particle quasimomentum  and $j=1,2,...,L$ \cite{Kolovsky2016}. The only other symmetry is the parity, which is defined as

\begin{equation}
    \hat{\mathcal{P}}\ket{n_1,n_2,...,n_{L-1},n_L} = \ket{n_L,n_{L-1},...,n_2,n_1}.
\end{equation}

\begin{figure}
\includegraphics[scale=0.5]{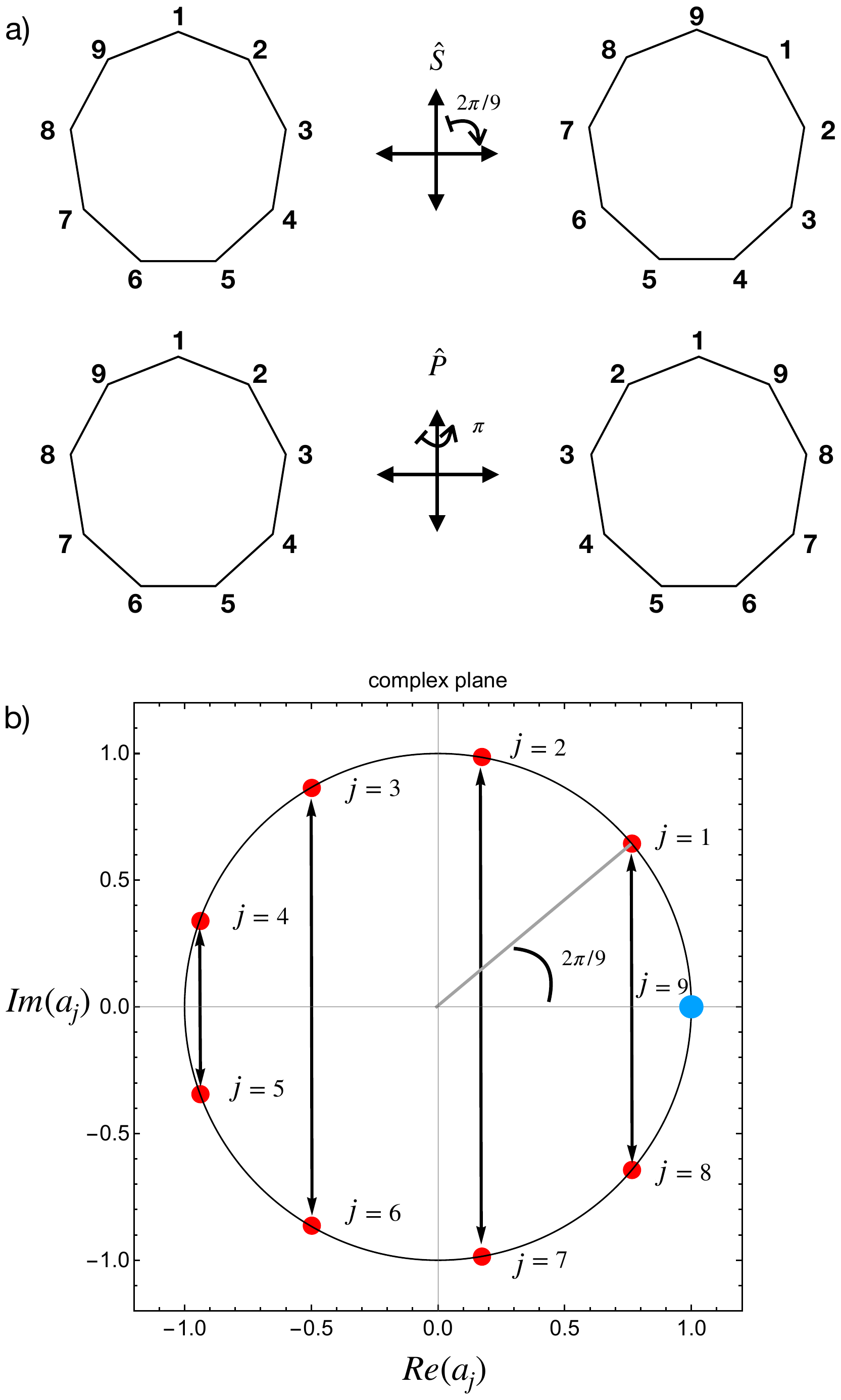}
\caption{a)  Schematic representations of the shift  $\hat{S}$ and parity  $\hat{\mathcal{P}}$ operators.  These operators describe  the symmetries of the BH model with periodic boundary conditions and  are the same as those of a regular polygon of $L$ sides. b)  Eigenvalues of $\hat{S}$ in the complex plane, for $L=9$ sites, indicated by  red dots. Pairwise subspaces related by complex conjugate eigenvalues of $\hat{S}$ have the same spectrum, these are linked by the black arrows. The subspace $j=9$ is the only non degenerate subspace and it has two parity symmetric invariant subspaces.}
 \label{fig:simetries}
\end{figure}

The shift and the parity operators commutes with the BH Hamiltonian \eqref{eq:Hamiltonian}. However $[\hat{S}, \hat{\mathcal{P}}]\neq 0$, 
which entails a pairwise degenerate spectrum between  states belonging to the subspaces with  eigenvalues $a_j$ and $a_j^*=a_{L-j}$. For   $a_{j=L}=1$ or $a_{j=L/2}=-1$ (the latter appearing only  for $L$ even)  the subspaces can be decomposed in two additional  subspaces with definite parity.  In this way we can classify the entire Hamiltonian spectrum. 

Fig. \ref{fig:simetries} a) shows a schematic representation of the shift and  parity operator acting on a  system of $L=9$ sites. From now on, we consider this number of sites and the same number of bosons $N=9$. For this choice,  the Hilbert space dimension is $\mathcal{D}=24310$, and it can be decomposed into $9$ subspaces associated with the eigenvalues $a_j$ of $\hat{S}$ as is shown in the Fig. \ref{fig:simetries} b). The dimensions of the subspaces are $\mathcal{D}_1 = \mathcal{D}_8=2700$, $\mathcal{D}_2 = \mathcal{D}_7=2700$, $\mathcal{D}_3 = \mathcal{D}_6=2703$ , $\mathcal{D}_4 = \mathcal{D}_5=2700$, $\mathcal{D}_9 = 2704$. Additionally, the subspace associated with $j=9$ can be decomposed in two subspaces with definite parity, whose dimensions are $\mathcal{D}_{9-even} = 1387$ and $\mathcal{D}_{9-odd} = 1317$. In Fig. \ref{fig:simetries} b) each subspace pair related by the black arrow has the same spectrum. Only the 9-subspace has no degeneracies, and the two subspaces with different parity have different spectrum. This is the full decomposition of the system in symmetries.
 
 For the Hamiltonian parameters,  we use the parametrization introduced in \cite{Kolovsky(2003)} $U= u$ and $J= 1-u$ with $u\in[0,1]$. With this parametrization the system is integrable in the two limits $u=0$ and $u=1$. Except for some few indicated cases, in the following sections we use  the  value  $u=0.5$ as a representative chaotic example. 

\subsection{Level statistics and density of states}

\begin{figure}
    \centering
    \includegraphics[scale=0.79]{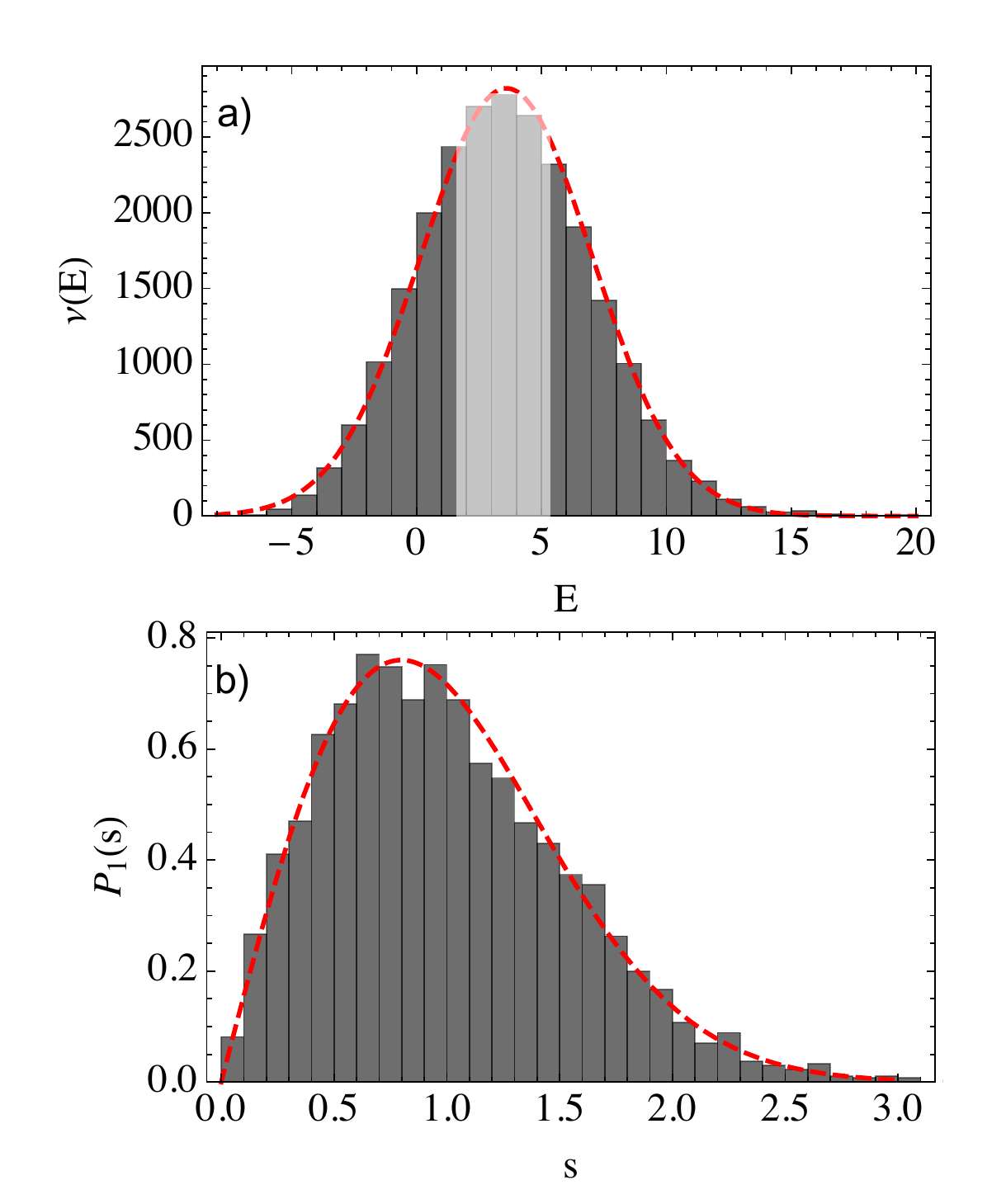}
    \caption{a) Gray bars depict the Density of states (DoS) obtained numerically for the BHM with $L=9$ sites and $N=9$ bosons.  A  Gaussian fit  is shown by  the red dashed line. The light-gray zone in the DoS represents  the energy interval used in panel (b) and  Figs.\ref{fig:Space_dsitribution}, \ref{fig:Survivals} and  \ref{fig3}(top). b) Nearest neighbour level spacing distribution for the unfolded spectrum  of the subspace associated to $j=1$  ($\kappa_1 = 2\pi/9$) symmetry subspace. The distribution coincides with the Wigner-Dyson surmise of the GOE ensemble (dashed red line). 
    \label{fig:Density_States}}
\end{figure}

We obtained the full spectrum of the model by exact numerical  diagonalization. The density of states (DoS) $\nu(E)$ is shown in the Fig. \ref{fig:Density_States}~a).  The DoS has a Gaussian form, consistent with the  finite size of the system. Due to the similar dimension of the subspaces the density of states in each one is very close to a ninth of the total density of states $\nu_{j}(E)=\nu (E)/9$, for $j=1,...,9$.  The light gray zone in Fig. \ref{fig:Density_States}~a)  depicts the  energy interval that we consider in the numerical calculations presented below for the spectral distributions and the survival probability. 
 This is a window of width equal to four (energy units) with center in the maximum of the distribution.

The nearest-neighbor spacing distribution for the unfolded energy levels of the 1-subspace, containing the central 80 \% of its spectrum,
 is shown  in Fig. \ref{fig:Density_States}~b). The red-dashed line is the Wigner-Dyson distribution for the Gaussian orthogonal  ensemble (GOE).  We have verified  the same  very good match with  the Wigner-Dyson distribution for the energy spacing of the other symmetry subspaces. 

\begin{figure}
\includegraphics[scale=0.68]{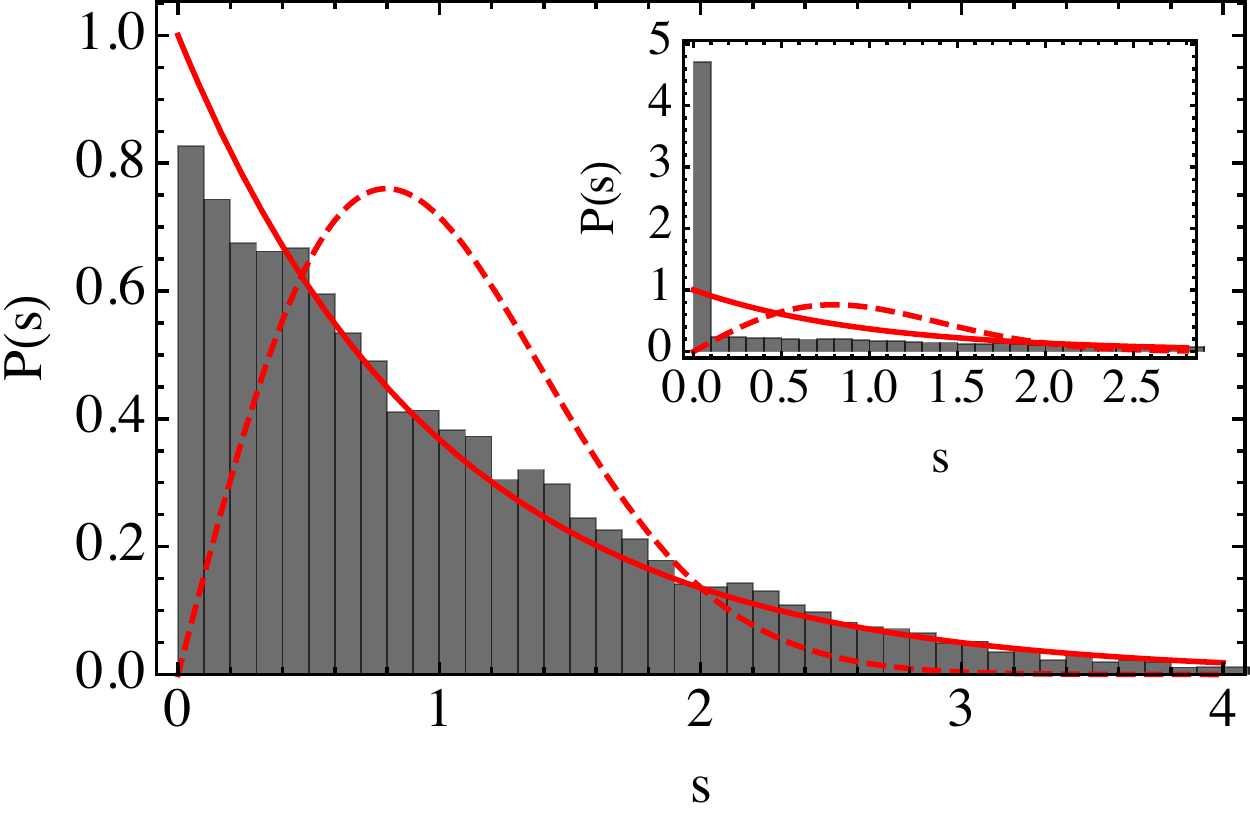}
\caption {Main panel shows the nearest neighbor spacing distribution of energy levels in the light gray region of Fig.\ref{fig:Density_States} considering  subspaces $j=1-4$ and $j=9$.  Inset shows the same distribution for the whole set of symmetry subspaces.}
\label{fig:Space_dsitribution}
\end{figure}

If we consider the  energy levels of all the symmetry sectors contained in  the same energy interval as before, we obtain the unfolded energy spacing distribution displayed   in the inset of  Fig.  \ref{fig:Space_dsitribution}. The distribution  shows a peak at zero energy, which comes from the exact respective  degeneracies between subspaces 1-4 and 8-5. 
If we remove these exact degeneracies by considering only subspaces 1-4 and 9, we get rid of the peak and obtain a distribution very close to a Poisson distribution of uncorrelated levels, as can be seen in the main panel of Fig.  \ref{fig:Space_dsitribution}. This well known result shows that the nearest neighbour distribution is unable to capture the intra-correlations between levels of the same symmetry  sectors, which become  hidden by the lack of correlations between levels of different symmetries.    

As we  show below, this is not the case of the survival probability, which is sensitive to the correlations in the spectrum even if several symmetry sectors are considered together.  In what  follows we discuss how  these correlations manifests as a dip in the temporal evolution of the survival probability, known as correlation hole, and compare it with  the spectral analysis performed above.

\section{\label{sec:survivalprobability}The Survival probability and the correlation hole}

The survival probability is a dynamical observable defined as the probability to find a given initial state $\ket{\Psi(0)}$ at time $t$,
\begin{equation}
    S_P(t) = \left|\braket{\Psi(0)|\Psi(t)}\right|^2.
\label{eq:SP1}    
\end{equation}
If we expand the initial state  $\ket{\Psi(0)}$ in the energy eigenbasis $\{\ket{\phi_k}\}$,  
\begin{equation}
\ket{\Psi(0)} = \sum_{k=1}^{\mathcal{D}}c_k\ket{\phi_k} ,
\end{equation}
where $\hat{H}_{BH}\ket{\phi_k}=E_k\ket{\phi_k}$ and $c_k = \braket{\phi_k|\Psi(0)}$, the survival probability reads
\begin{equation}
    S_P(t) = \left|\sum_k |c_k|^2 e^{-iE_kt} \right|^2.
\end{equation}

\subsection{Initial decay and asymptotic value}
The survival probability can be expressed as the squared modulus of the Fourier transform of the local density of states (LDOS)
\begin{equation}
    S_P(t) = \left|\int \mathcal{G}(E) e^{-iEt}dE\right|^2.
\end{equation}
where the LDOS,  $\mathcal{G}(E) = \sum_k |c_k|^2 \delta(E-E_k)$, is the energy distribution weighted by the components of the initial state.
By considering a smoothed approximation to the LDOS,  $\rho(E)\approx \mathcal{G}(E)$, we can obtain an analytical expression for the initial decay of $S_P(t)$  \cite{Torres-Herrera(2017),Lerma-Hernandez(2019)}. For instance, for a rectangular smoothed profile
\begin{equation}
    \rho(E) = \left\{\begin{array}{lr}\frac{1}{2\sigma_R} & \text{for } E\in[E_c-\sigma_R,E_c+\sigma_R] \\
                                        0 & \text{otherwise}
                                        \end{array}\right. ,
\end{equation}
we obtain a $sinc$ squared  function for the  initial decay of $S_P(t)$
\begin{equation}
    S_p^{bc}(t) = \frac{\sin^2(\sigma_R t)}{(\sigma_R t)^2},
    \label{eq:Sp_bc}
\end{equation} 
where the super-index $bc$ indicates that this expression is valid before the dynamics is able to resolve the correlations in the spectrum. 
In the following, we consider  a such  rectangular energy profile with parameters determined by the energy window of  Fig. \ref{fig:Density_States}, 
 where $E_c = 3.60$ is the center of the distribution and $\sigma_R=2$ is its half-width.

The initial decay holds up to a temporal scale where the dynamics is able to resolve the discrete nature of the energy spectrum. For $t\rightarrow\infty$, the survival probability fluctuates around an asymptotic value, $S_P^\infty$, which can be determined as follows. By expanding  the squared modulus  in Eq.(\ref{eq:SP1}), we obtain 
\begin{equation}
    S_P(t)=\sum_{k\neq l}|c_l|^2|c_k|^2e^{-i(E_k-E_l)t}+ \sum_{k}|c_k|^4 .
\label{eq:SP2}
\end{equation}
The asymptotic value can be obtained by considering a temporal average of this expression
\begin{equation}
 S_P^\infty =  \lim_{t\to\infty} \frac{1}{t}\int_0^{\infty}S_P(t')dt'.
\end{equation}
In the absence of degeneracies, the first term in the RHS of Eq.({\ref{eq:SP2}}) cancels out in average and 
\begin{equation}
   S_P^\infty =  \sum_{k}|c_k|^4,
   \label{eq:spinfty}
\end{equation}
which is the case of the BHM  when only one symmetry sector is considered.  In the case of degeneracies, that appear  in the  BHM  when several symmetry sectors are considered, the first term in the RHS of  Eq.(\ref{eq:SP2}) contributes with extra terms to the asymptotic value, which is now given by
\begin{equation}
   S_P^\infty =    \sum_{E_k}\left(\sum_{m=1}^{d_{E_k}}\left |c_{E_k, m}\right |^2\right)^2,
   \label{eq:spinfty2}
\end{equation}
where $c_{E_k,m }$ is the  component of the initial state in the energy level  $|E_k, m\rangle$   with degeneracy $d_{E_k}$, 
$$
c_{E_{k},m}=\langle E_k;m|\Psi(0)\rangle \ \ \ \ (m=1,..., d_{E_k}). 
$$
 
\subsection{Initial states}

In between the initial decay and the saturation of the dynamics, correlations in the energy spectrum manifest as a dip of the $S_P$ below the asymptotic value. To reveal the presence of this correlation hole, averages over different initial states have to be considered \cite{Torres-Herrera(2017), Torres-Herrera(2017-2),Torres(2018),Torres-Herrera(2019), Schiulaz(2019),Lerma-Hernandez(2019)}. In this paper we consider  ensembles of initial states with components different to zero only in the energy interval $[E_c-\sigma_R,E_c+\sigma_R]$ and whose squared modulus are randomly chosen as follows  
\begin{equation}
    |c_k|^2 = \frac{r_k f(E_k)}{\sum_q r_q f(E_q)}, 
 \label{ck2}     
\end{equation}
where $r_k\in[0,1]$ are random numbers coming from a  uniform distribution. The function $f(E)= \rho(E)/\nu(E)$ guarantees that   the random  initial state has the selected energy  profile $\rho(E)$. This is achieved by  compensating, with the denominator,  for changes in the density of states.

\subsection{Analytical expressions for the survival probability }

\begin{figure*}
    \centering
    \includegraphics[scale=0.8]{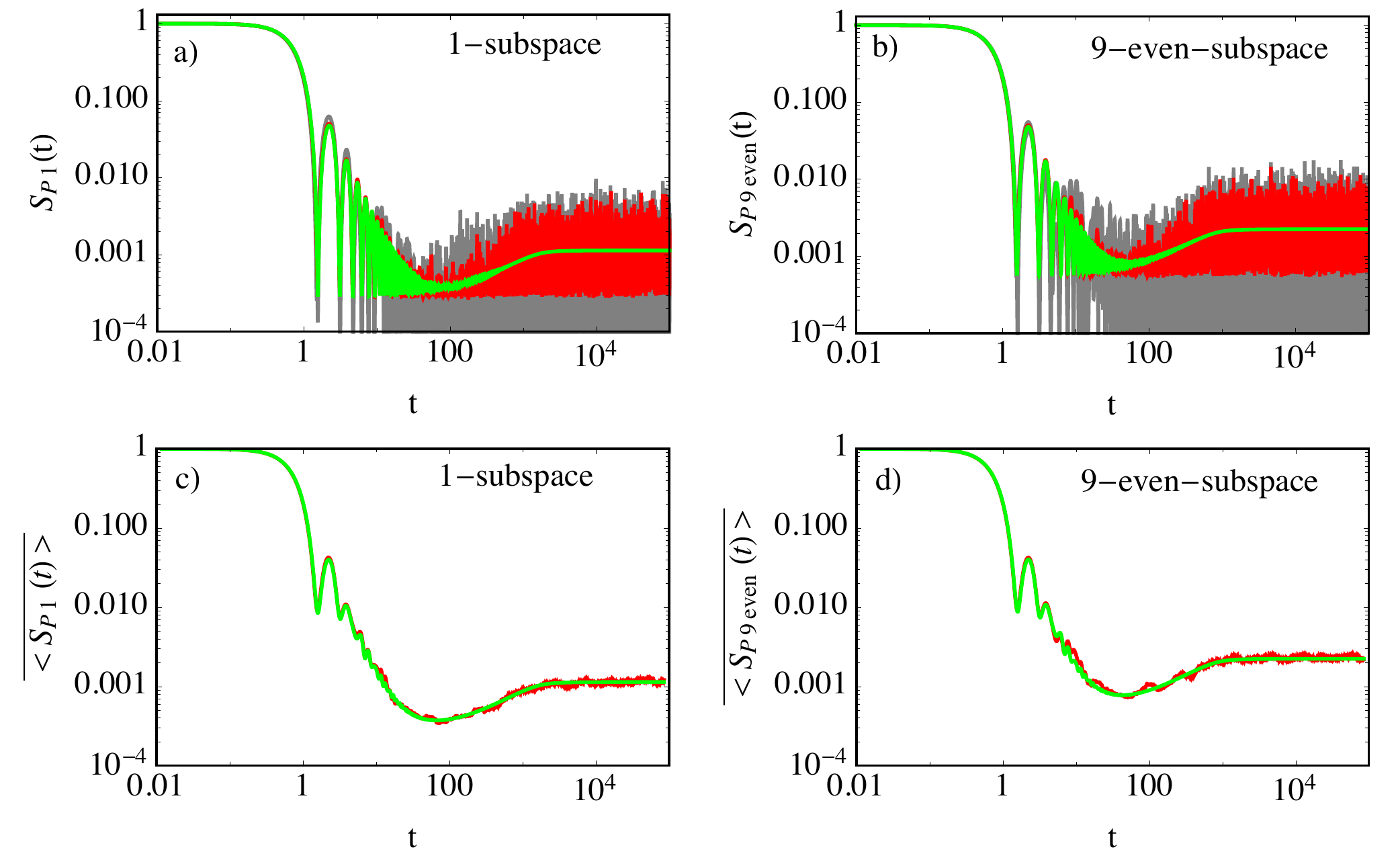}
    \caption{Survival probability as a function of time in log-log scale. Panels a) and b) show, for subspace $j=1$ and $j=9$-even respectively,  the survival probability for individual members of the ensemble (gray lines),  dark red lines represent the ensemble averages and light green lines are obtained from the analytical expression \eqref{eq:S_p-analitical}.  A similar  behavior (not shown)  was found for the rest of symmetry subspaces.   Panel c) and d) show    averages over temporal windows of constant size in log scale plotted versus the mean time  of the respective temporal windows  for  the numerical  ensemble average of $S_P(t)$ (dark red) and for the analytical expression  of the survival probability (light green).  Panel c) is for the 1-subspace and panel d) for the  9-even-subspace.}
    \label{fig:Survivals}
\end{figure*}
\subsubsection{One symmetry sector}

In \cite{Lerma-Hernandez(2019)}, by following an idea initially introduced in \cite{Herrera(2018)}, an analytical expression for the ensemble average of the survival probability was  derived, which applies for the case of one  sequence of non-degenerate energy levels with  energy density $\nu$ and  correlations similar to those of random matrices of a Gaussian orthogonal ensemble (GOE)   
\begin{equation}
    \braket{S_P(t)} = \frac{1-\langle S_P^\infty\rangle}{\eta -1}\left[\eta S_P^{bc}(t) - b_2\left(\frac{t}{2 \pi \bar\nu}\right)\right] + \langle S_P^\infty\rangle,
    \label{eq:S_p-analitical}
\end{equation}
here $S_P^{bc}(t)$ is given in Eq.(\ref{eq:Sp_bc}), $\bar{\nu}$ is the mean DoS in the energy interval,  and   $\eta$ is the effective dimension of energy levels available for  the ensemble, which  is given by
\begin{equation}
\eta=\frac{1}{\int dE \frac{\rho^2(E)}{\nu(E)}}=\frac{4 \sigma_R^2}{\int_{E_c-\sigma_R}^{E_c+\sigma_R}dE \frac{1}{\nu(E)}},
\label{eq:eta}
\end{equation}
where in the last equality we have used the rectangular  profile for the energy distribution. When the energy interval is approximately centered in the middle of the Gaussian distribution for $\nu(E)$,  the last equation can be approximated by substituting $\nu(E)$ in the denominator by its average in the energy interval ($\bar{\nu}$), leading to a simple expression    $$\eta=2\sigma_r \bar{\nu},$$
which  equals  the number of states in the energy interval.
  The asymptotic value $\braket{S_P^\infty}$ is obtained by averaging expression (\ref{eq:spinfty}), which can be shown \cite{Lerma-Hernandez(2019)} to be given by   
\begin{equation}
\braket{S_P^\infty} = \frac{\braket{r^2}}{\braket{r}^2}\frac{1}{\eta}= \frac{4}{3\eta},
\label{eq:SPinfan1}
\end{equation}
where $\braket{r^n}$ is the n-th moment of the  distribution of the random numbers  and in the last equality we have used that for an uniform distribution $\langle r^n\rangle=1/(n+1)$.

The second term inside the brackets in Eq. \eqref{eq:S_p-analitical} is the two-level form factor of the GOE ensemble \citep{Alhassid(1992)},
\begin{equation}
    \begin{split}
        b_2(t) &= \left[1-2t+t\ln(2t + 1)\right]\Theta(1-t)\\
        &+ \left[t\ln\left(\frac{2t+1}{2t-1}\right)-1\right]\Theta(t-1),
    \end{split}
\end{equation}
where $\Theta$ is the Heaviside step function. The two-level form factor brings the survival probability from its minimum value up to the asymptotic value $\braket{S_P^\infty}$, creating the dip that is known as the {\em correlation hole}. Although the averaged survival probability can display oscillations, the presence of a hole which can be described by \eqref{eq:S_p-analitical} is a direct signature of the existence of correlated eigenvalues, and it does not develop in systems with uncorrelated eigenvalues. 

\subsubsection{Whole set of  symmetry sectors}
The formula \eqref{eq:S_p-analitical} is applicable to the BHM when only one symmetry sector is considered.  Here we extend  Eq.\eqref{eq:S_p-analitical} to the case where   the whole set of symmetry  sectors are included.  The general formulae and  details of the derivation can be seen in the Appendix. Here we present the results for random numbers coming from an uniform distribution,  and for the BHM with  $L=9$ sites and  $N_S=10$ symmetry sectors whose densities are  approximately $\nu_j=\nu/9$ for $j=1,...,8$ and $\nu_{9-even}=\nu_{9-odd}=\nu/18$, while  their  degeneracies are   $d_j=2$ for $i=1,...,8$ and $d_{9-even}=d_{9-odd}=1$. For this particular case the ensemble average of  the survival probability reads
\begin{align}
\langle S_P(t)\rangle_{_{\hbox{\scriptsize $a$}}} =&\frac{ 1-\langle S_P^\infty\rangle_{_{\hbox{\scriptsize $a$}}} }{\eta-1}
\left[ \eta S_P^{bc}(t)-\frac{16 b_2\left(\frac{9t}{2\pi \bar{\nu}}\right)+b_2\left(\frac{18 t}{2\pi \bar{\nu}}\right)}{9} \right] \nonumber\\
 &+\langle S_P^\infty\rangle_{_{\hbox{\scriptsize $a$}}},
 \label{eq:span2}
\end{align}
where the density $\nu(E)$ that has to be considered in  Eq.(\ref{eq:eta}) to calculate $\eta$ is the density of the whole spectrum.    The asymptotic value of $\braket{S_P(t)}_a$ is  now  obtained by ensemble averaging  Eq.(\ref{eq:spinfty2}), which yields
\begin{equation}
\langle S_P^\infty\rangle_{_{\hbox{\scriptsize $a$}}} =  \frac{4}{3\eta}+\frac{8}{9}\frac{\left(1-\frac{4}{3\eta} \right)}{\eta-1}\xrightarrow{\eta\gg 1} \frac{20}{9 \eta }.
\label{eq:spasyman2}
\end{equation}
The key point is that, again, as in  the case of only one symmetry sector, the intra-correlations of levels  in the same symmetry sectors,  brings the survival from its minimum up to the asymptotic value, creating a correlation hole.

At variance with the one-symmetry case, in this case the correlation hole is  governed  by a pair of  two-level form-factors  $b_2$. One coming from the subspaces 1-8 which implies that the density entering in its argument is $\bar{\nu}_j=\bar{\nu}/9$, while the second comes from intra-correlations in the spectrum of subspaces $9$-even and $9$-odd. That is why in the argument of this $b_2$ function enters $\bar{\nu}_{9-\text{even}}=\bar{\nu}_{9-\text{odd}}=\bar{\nu}/18$.   
Since the dynamics of subspaces 1-8 have the same temporal scale,  their individual contributions add coherently to build up the correlation hole and they dominate the second term inside the parenthesis in (\ref{eq:span2}).  

\section{\label{sec:results}Numerical vs analytical  results}

In this section we compare the analytical expressions for the survival probability from the previous section with numerical results obtained by diagonalizing numerically the BHM for one and the whole set of symmetry sectors. We also study the way as the signatures of  chaos, i.e. the correlation hole, dilutes as we approach an integrable limit. 

\subsection{Correlation hole for an ensemble of initial states in the same symmetry subspace}

In Fig. \ref{fig:Survivals} we compare numerical results for the survival probability with the analytical expression given by Eq. \eqref{eq:S_p-analitical}.  We employ the same distribution of random initial states in the energy interval $[E_c - \sigma, E_c + \sigma]$ indicated by light-gray columns at the center of the DoS in Figure \ref{fig:Density_States}. For each $\kappa_j$ subspace, we consider as many random initial states as the number of energy levels of that subspace in the energy interval, i.e,  $\sim 1170$ for the $j=1-4$-subspaces and  $\sim 590$ initial states for the subspaces  9-even and  9-odd. The exact numbers are shown in the third column of Table \ref{tab:Comparation}.

In  Fig. \ref{fig:Survivals}~(a) and (b), the gray lines show the survival probability as a function of time for  different random initial states with components in the indicated $\kappa_j$-subspace, the red lines are the numerical averages over the ensemble, whereas   the green line is the analytic expression given in Eq. \eqref{eq:S_p-analitical} with parameters determined from Eqs.(\ref{eq:eta}) and (\ref{eq:SPinfan1}). 
 We can see that the analytical expression properly describes the temporal trend of the numerical ensemble average. At short times and before the $S_P(t)$ attains its minimum value, the fluctuations of the numerical ensemble average are small, which is a consequence of the fact that the  initial decay is determined entirely by the smoothed energy profile of the initial states, Eq. \eqref{eq:Sp_bc}, which is the same for every member of the ensemble. At the temporal scale of the correlation hole and beyond, the fluctuations in the numerical ensemble average are relatively larger. 
 To further reduce these fluctuations we consider  averages over temporal windows of constant size in log scale. By considering the same smoothing procedure in the analytical expression, we obtain Figures  \ref{fig:Survivals}~(c) and (d), which show an excellent agreement between the analytical expression and numerical averages, and make evident the correlation hole, confirming the existence of a GOE  correlated spectrum and thus quantum chaos in this energy region of the Bose-Hubbard model.   
 
\subsubsection{\label{sec:level11}Survival probability for different regions of the spectrum}
\begin{figure}
\begin{tabular}{cc}
\rotatebox{90}{ \hspace{100pt} {\hbox{\Large{$\overline{\braket{S_{P}(t)}}$}}}} &\includegraphics[scale=0.5]{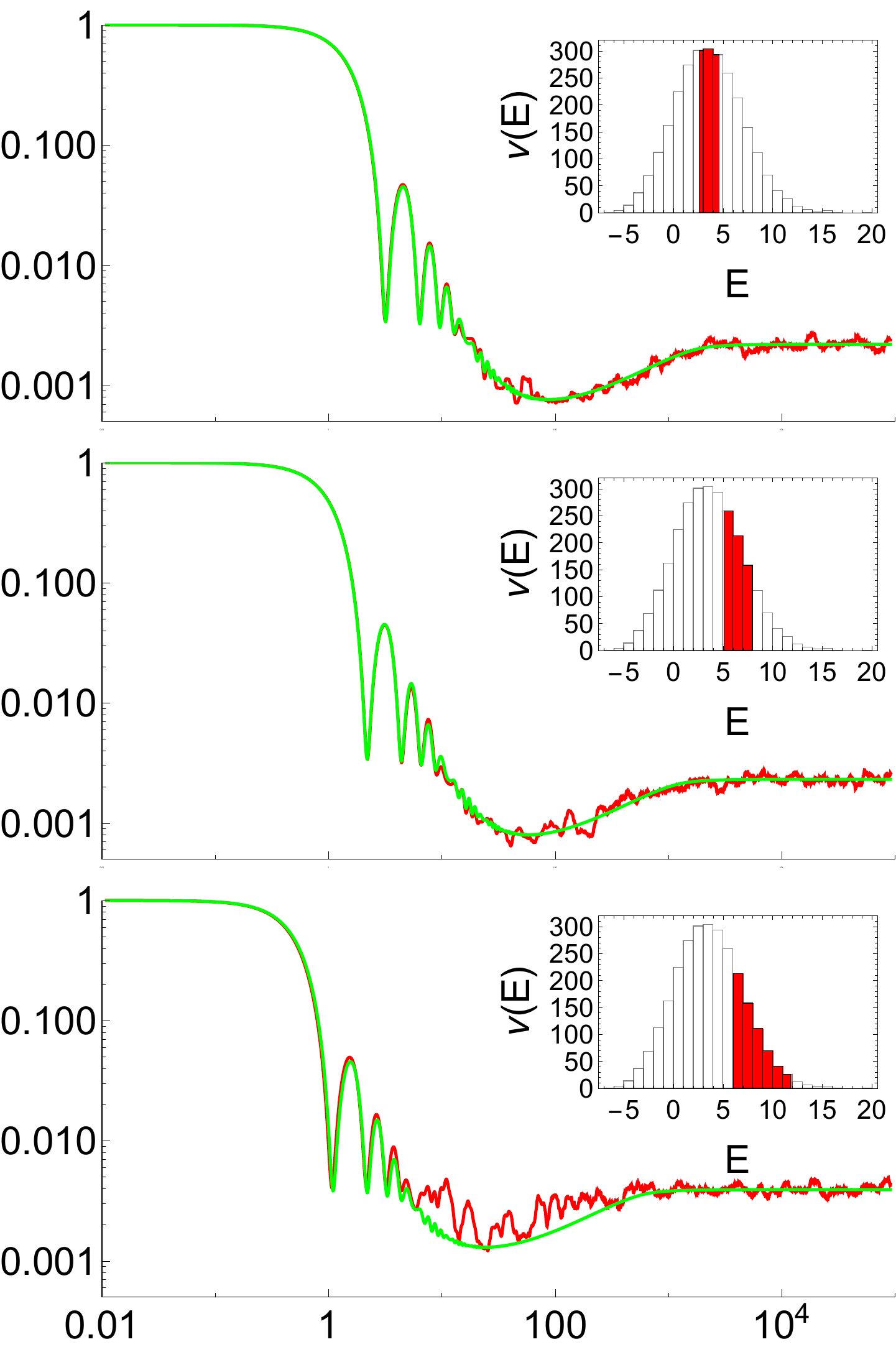}\\
& {\hbox{\Large{$t$}}}
\end{tabular}
\caption{Averages over temporal window of constant size in log scale of the numerical ensemble average of the Survival probability, plotted versus the mean value of the respective temporal windows (red dark line). Three ensembles over energy regions with $600$ levels each were considered. The energy regions of each ensemble are indicated in the histograms shown as insets.  Light green lines indicate the same temporal averages of the analytical expression \eqref{eq:S_p-analitical}. Only energy levels of  the $j=1$-subspace were considered. From top to bottom the energy regions move from the center of the spectrum to its  highest border.   }
\label{Fig:Sp_regiones}
\end{figure}

In the previous section we have analyzed  initial states with random components in the central region of the spectrum. Now we study
what happens with the survival probability if we select  initial states in energy regions approaching  the border of the spectrum. It is known that the universal statistical properties of the chaotic spectra are not applicable to the borders of the spectra, which are model dependent\cite{Gutzwiller(2013)}. This is confirmed in the behavior of the   survival probability shown in Fig.\ref{Fig:Sp_regiones} with dark red lines. Results for three ensembles are shown, one ensemble, as before and used as reference, is located in  the center of the DoS and the other two approach the high energy border of the spectrum.
We consider again  rectangular energy profiles, and  move the energy window to the large energy regions using eigenstates of the 1-subspace. The energy windows we consider are shown in the insets of  Fig. \ref{Fig:Sp_regiones}.  The number of energy levels contained in the  three energy windows  is equal to $600$.         

We observe, as in  Fig.\ref{fig:Survivals}, a correlation hole for the first and second ensemble of states,  located respectively  in the central part of the spectrum and  in a region with slightly higher energies. In both cases the survival probability and correlation hole are  very well described by the analytical expression (\ref{eq:S_p-analitical}), shown with light green  lines.  Since these two ensembles prove energy levels far enough of the borders of the spectrum, the universal behavior expected from the GOE ensemble is clearly observed. 

For the third ensemble, located  close to the border of the spectrum, we can observe a clear  deviation respect to the universal behavior  indicated by the light green  line obtained from Eq.(\ref{eq:S_p-analitical}).  We observe a correlation hole in the numerical results that is smaller than the one coming from the analytical expression. This implies that, contrary to the two previous ensembles, only a fraction of levels in this  energy region have GOE correlations. The participation, in this third ensemble, of  energy levels  located  in the higher part of the spectrum, not only diminish the depth of the correlation hole, but also produce larger oscillations in the survival probability, as  can be observed in the bottom panel of Fig.\ref{fig:Survivals}.

\subsubsection{\label{sec:level12}Survival probability for different Hamiltonian parameters}

\begin{figure*}
	\centering
	\includegraphics[scale=0.52]{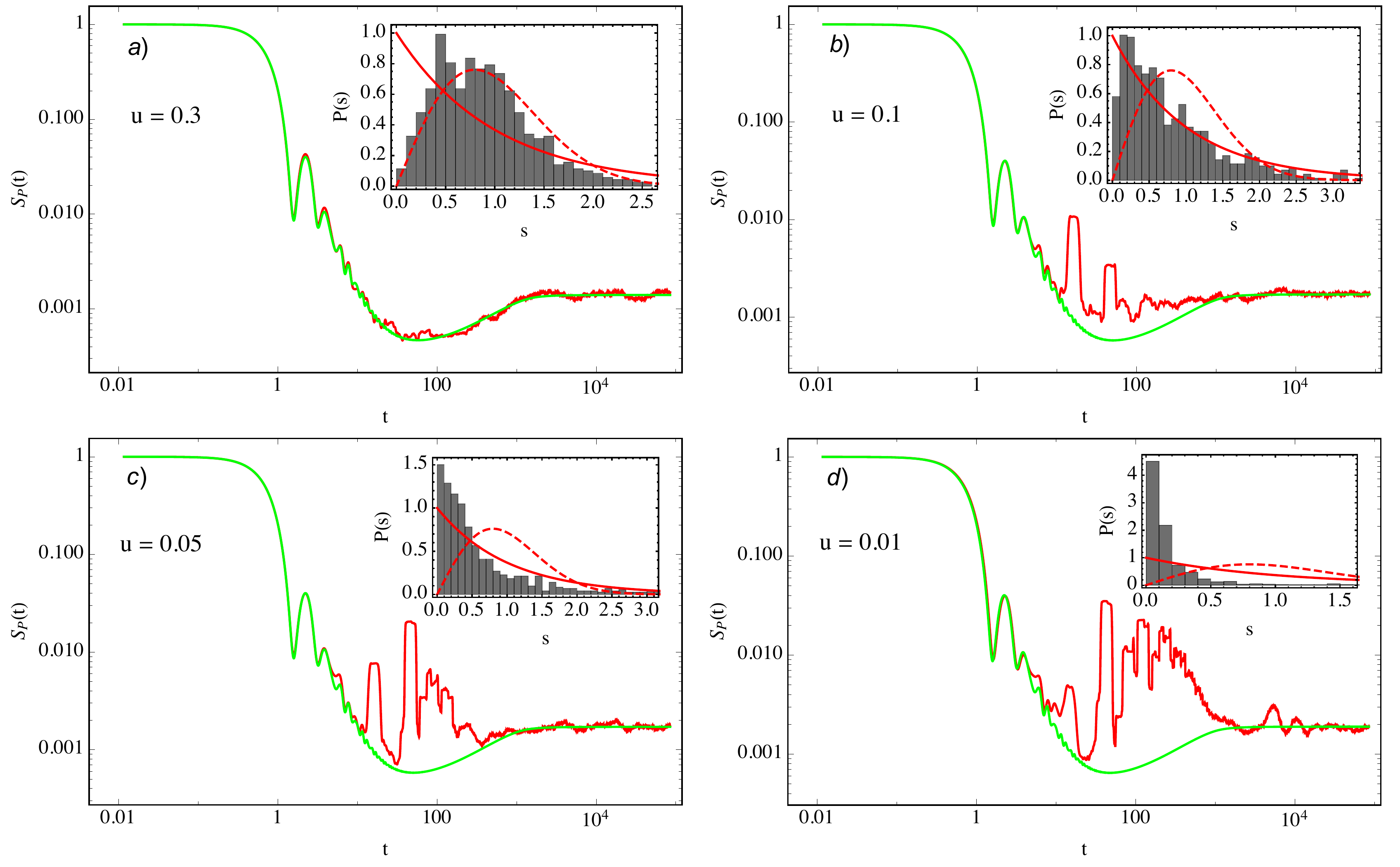}
	\caption{Averages over temporal windows of the survival probability for  initial states with components in the central part of the spectrum of the 1-subspace. The dark red curve is the numerical ensemble average of the survival probability, while  the light green line is the analytical expression \eqref{eq:S_p-analitical}. Four different  couplings $u$, indicated in each panel,  were used. The insets display the corresponding nearest neighbor spacing distribution, where the continuous red line is  the Poisson distribution and the dashed line is the Wigner-Dyson distribution.}
	\label{Fig:Sp_Variando_u}
\end{figure*}

Above we have analyzed the properties of the spectrum of the BHM in the chaotic regime with parameter $u=0.5$. In this section we study the survival probability with initial states in the same central region of the spectrum of  the symmetry sector $\kappa_1$, but now for   values of the parameter $u$ approaching the integrable limit $u=0$.

In Fig \ref{Fig:Sp_Variando_u} we show the combined ensemble and temporal average of the survival probability, both numerical (red) and analytical (green), for four different values of the $u$ parameter. For $u=0.3$  Fig. \ref{Fig:Sp_Variando_u} a) shows a perfect match between numerical and analytical results and a clear presence of the correlation hole. This is consistent with the nearest neighbor energy distribution shown in the inset, which is very well described by the Wigner-Dyson surmise. As we approach the integrable limit,  the nearest neighbor energy differences histograms in the insets of Figs.(b,c,d) show that the spectrum correlations disappear. Accordingly,  except for the initial decay and asymptotic value, the analytical expression no longer describes the numerical survival probability and  instead  of  a correlation hole we observe revivals whose amplitude increases as the value of $u$ approaches the integrable limit.
   
\subsection{Correlation hole for an  ensemble of initial states in the full space}
\begin{figure}[h!]
    \centering
    \includegraphics[scale=0.7]{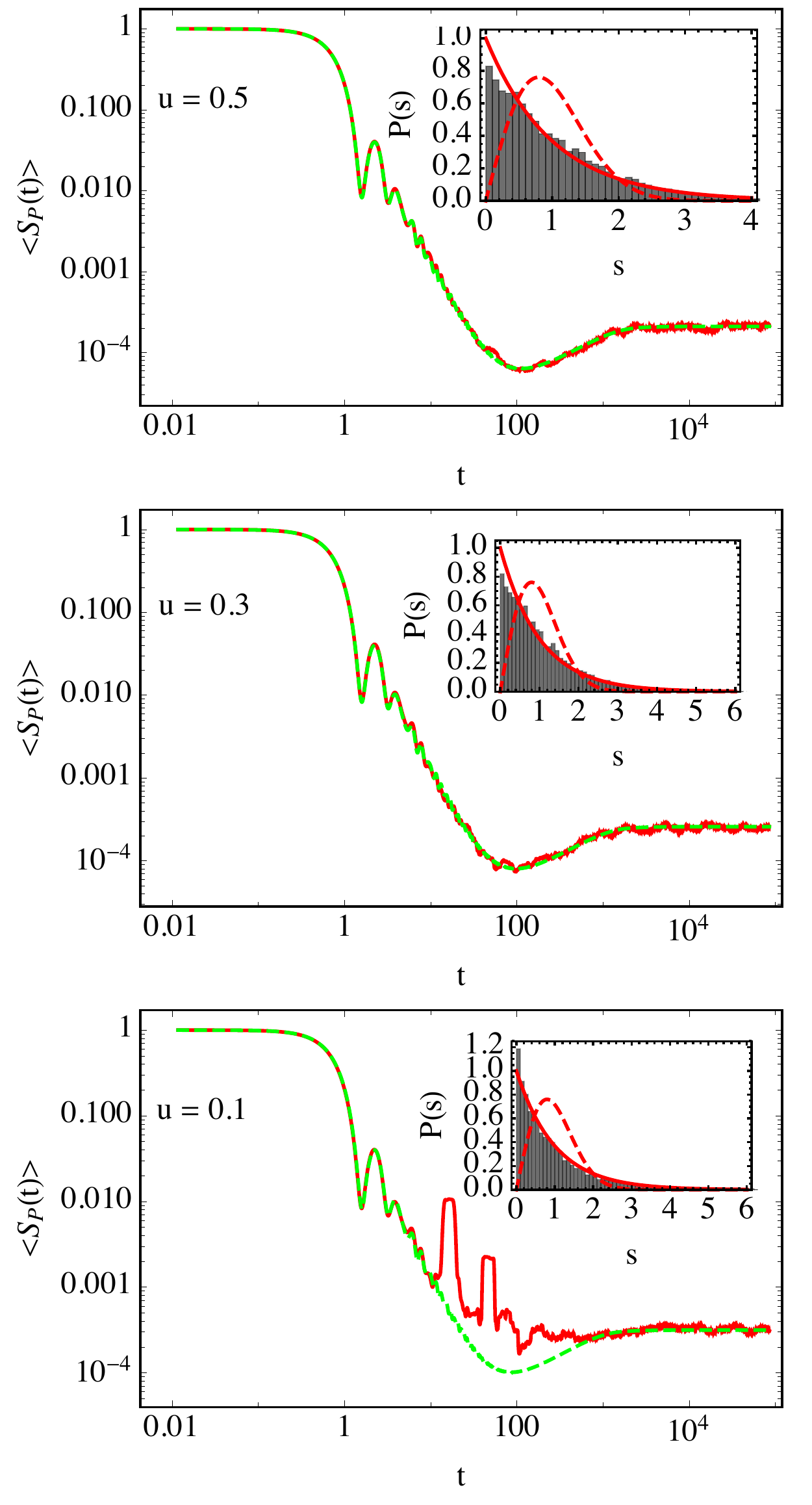}
    \caption{Dark red lines show the temporal mean of  the survival probability for the ensemble average  of  10581 initial states  with the same number of components  inside the central energy interval of  all the  symmetry  subspaces.  Three different couplings $u$, indicated in each panel, were employed. For the top panel the energy interval used is  indicated by light-gray  bars in panel a) of  Fig. \ref{fig:Density_States}. Light green lines are the temporal mean of the analytical expression in Eq.\eqref{eq:span2}. Insets  show the corresponding nearest neighbor spacing distribution of energy levels considering only   subspaces $j=1-4$ and $j=9$.}
    \label{fig3}
\end{figure}
In this section we present numerical and analytical results for random initial states but considering now the whole set of symmetry sectors.
The combined temporal and ensemble average of  $S_P(t)$  for a such  set  of initial states  with the participation of levels in the energy interval indicated by light-gray bars in Fig.\ref{fig:Density_States} is shown with a red line in Fig \ref{fig3}(top). 
The important finding is that, even if we employ the complete energy spectrum without any consideration about symmetries, the survival probability exhibits a clearly visible correlation hole, which is again very well described  by the analytical expression (green line), now given by Eq. \eqref{eq:span2}, whose  parameters  are determined  from  Eqs. \eqref{eq:eta} and (\ref{eq:spasyman2}). 

 By inspecting Eq.\eqref{eq:span2}, we observe that the correlation hole is governed mainly by the first term with the two-level form factor $b_2$ coming from the subspaces $\kappa_j=\kappa_1,...,\kappa_8$, which is $16$ times larger than the second term with the  $b_2$ function coming from the subspaces $9$-even and $9$-odd. The presence of $(L-2)/2$ (for even $L$) or $(L-1)/2$ (for odd $L$) sequences of energy levels with similar densities and GOE intra-correlations,   is the general scenario that can be found in the BHM for an arbitrary  number of sites, and therefore no-exception can  be foreseen for the appearance of the correlation hole in the chaotic regime of the BHM for arbitrary size. To illustrate the generality of the previous scenario, we show in Fig. \ref{fig3}(middle) the same analysis as before, but now for a coupling $u=0.3$. We observe a perfect match between analytical and numerical results, allowing to conclude that this coupling corresponds also to a chaotic regime.
 
  Finally, in the bottom panel of Fig.\ref{fig3} we present the survival probability  for a coupling, $u=0.1$, close to the integrable limit. Similarly  to the results shown in Fig.\ref{Fig:Sp_Variando_u}(b,c,d),  only  the initial decay and asymptotic value are well described by the analytical expression. At the temporal scale where the correlation hole would develop, the numerical results show instead partial revivals, indicating that for this coupling there are no GOE correlations even for states belonging to  the same symmetry sector. 

Fig.\ref{fig3} shows that for an spectrum coming from different symmetry sectors,  the correlation hole in the survival probability serves as a good indicator of quantum chaos, differently to the nearest-neighbour-energy differences distribution, which, as shown in the insets of Fig.\ref{fig3}, are  Poisson-like  in the three cases and thus useless to distinguish a chaotic from  a regular regime.   
      
Even in more general chaotic cases, distinct to the BHM, the analytical general formula for the ensemble average of $S_P$ presented in the Appendix, shows that the correlation hole is a general feature appearing in the evolution of the $S_P(t)$  and can be used as a reliable indicator of quantum chaos without having to classify the energy levels according to their symmetries. 
 In this context, it is appropriate to mention a  recent  study of the temporal evolution of the Survival Probability  in the chaotic region of  the Dicke model \cite{Villasenor2020}, where  initial states  mixing  two subspaces with different parity symmetries are considered.   In that reference is also reported the need to employ the full effective dimension $\eta$,  but only the  density of states of every  subspace  to describe  analytically  the presence of the correlation hole in the numerical averages. This result is a particular case of the general formula described in the Appendix for the survival probability when several symmetry subspaces are included.     

\begin{table}
\centering
\begin{tabular}{|c|c|c|c|}\hline
$\kappa_j$-subspace &  $\braket{S_P^\infty} \times 10^3$& $\eta$ & $\overline{\nu}$\\ \hline \hline
1, 8 & 1.11 & 1174 & 293.5\\ \hline
2, 7 & 1.13 & 1175 & 293.7\\ \hline
3, 6 & 1.13 & 1178 & 294.5 \\ \hline
4, 5 & 1.13 & 1179 & 294.7\\ \hline
9 even & 2.23 & 597  & 149.25 \\ \hline
9 odd  & 2.32 & 574 &143.5\\ \hline
 Complete & 0.210 & 10581 & 2645.25\\ \hline
\end{tabular}
\caption{Parameters of the analytical equation  \eqref{eq:S_p-analitical} plotted in Fig.\ref{fig:Survivals} for the $j=1,9$-subspaces. Second and third column come, respectively,  from  equations   \eqref{eq:SPinfan1} and \eqref{eq:eta}. Last row shows the parameters used in the analytical expression   \eqref{eq:span2} plotted in the  top panel of Fig.\ref{fig3}. Second and third-column values come from eqs.\eqref{eq:spasyman2} and \eqref{eq:eta}. For the cases shown the parameter $\eta$ coincides with the respective number of states in the energy window shown by light gray columns in Fig.\ref{fig:Density_States}. 
\label{tab:Comparation}}
\end{table}

\section{\label{sec:conclusions} Conclusions}

In the present work we have studied dynamical signatures of quantum  chaos in the Bose-Hubbard model in a ring configuration. After exhibiting that,  when levels  in the same subspace of the shift symmetry are considered, the unfolded nearest neighbor distribution match the  Wigner-Dyson surmise of the Gaussian orthogonal ensemble; we have shown that this is not the case when the full spectrum is included: when the energy degeneracies are removed from the whole spectrum, the distribution becomes Poisson-like.  
Then, we have studied the ensemble average of the Survival Probabililty $\braket{S_p(t)}$ for initial states in each symmetry subspace and  located  in  different energy windows. We have shown that, when the energy window is located far from the border of the spectrum,  a correlation hole in the survival probability  develops in all the  subspaces. It was also shown that the complete evolution of the survival probability  is very well described by an analytical expression deduced employing Random Matrix Theory.  These results exhibit  the presence of the correlation hole in the Survival Probability as a clear signature of quantum chaos.

Unlike the well established spectral analysis, we found that the correlation hole is a signature of quantum chaos that  does not require the classification of states according to their symmetries. Contrary to the analysis of the nearest-neighbor-energy-differences, in the survival probability  the intra-correlations of levels in the same symmetry sectors are not washed out by the absence of correlations between levels coming from different symmetry sectors.  Analytical expressions supporting the previous result were given and shown to describe perfectly the numerical results. The existence of the correlation hole was also correlated with the presence of quantum chaos for different values of the interaction u.
    
The finding that the correlation hole is present even when no separation in symmetries is performed,  exhibits  the survival probability as a very useful and powerful tool to identify the presence of quantum chaos in systems where the symmetry separation is far from trivial, either in the theoretical analysis or in its experimental observation.
 
An interesting extension of the studies presented here would be to investigate the dynamics of  special sets of initial states, like Fock states. The dependence of the dynamics of the survival probability on the dimension of the systems and the study of other observables would be other interesting directions to be investigated.

\section{Acknowledgments}
We thank L. Santos and J. Torres for their useful comments. 
We acknowledge the support of the Computation Center-ICN, in particular to Enrique Palacios, Luciano D\'iaz and Eduardo Murrieta. Also we acknowledge funding from Mexican Conacyt project CB2015-01/255702, and DGAPA-UNAM projects No. IN109417 and IN104020.

\appendix

\section{Ensemble average of the Survival Probability for several  degenerate sequences of energy levels}
Here we generalize the analytical expression for the ensemble average of the Survival probability  in chaotic regimes derived in \cite{Lerma-Hernandez(2019)} to the case of several symmetry sectors with energy degeneracies, which is the general case of the Bose-Hubbard model.

Let $N_e$ be the number of energy sequences. We assume that the unfolded energies   in every sequence have the same correlations as the Gaussian Orthogonal Ensemble, and no correlations exist between energies of different sequences.
Let $d_i$, $\nu_i$ and $L_i$ be, respectively,   the degree of degeneracy of each energy level, the Density of States  and the number of energy levels in  the $i$-th energy sequence ($i=1,..., N_e$). The following relation holds $\sum_i^{N_e} d_i\nu_i=\nu$, where $\nu$ is the Density of States of the whole spectrum.
The survival probability can be expressed as
$$
SP(t)=\sum_{E_{i k}\not=E_{i' k'}}\left|c_{i k}^{(m)}\right|^2 \left|c_{i' k'}^{(m')}\right|^2 e^{-i(E_{i k}-E_{i'k'})t}+
$$
$$
\sum_{i=1}^{N_E}\sum_{k=1}^{L_i}\left( \sum_{m=1}^{d_i}\left |c_{i k}^{(m)}\right |^2\right)^2,
$$
where $c_{i k}^{(m)}$ are the energy components of the initial state $|\Psi_o\rangle=\sum c_{i,k}^{(m)}|E_{i,k};m\rangle $. The first term in the expression has an infinite temporal  average equal to  zero and the second term ($S_P^\infty$) gives the asymptotic value of the survival probability.  

As stated in the main text, the  components of the initial states are chosen randomly according to
$$
\left|c_{i k}^{(m)}\right |^2= \frac{r_{ik}^{(m)}f(E_{ik})}{\sum_{i'k'm'} r_{i'k'}^{(m')}f(E_{i'k'}) },
$$
where $f(E_{ik})=\rho(E_{ik})/\nu(E_{ik})$ and $r_{ik}^{(m)}$ are positive random numbers from a probability distribution $p(r)$.  

To derive an analytical expression for the ensemble average of the Survival probability, we proceed similarly as \cite{Lerma-Hernandez(2019)}, and consider the following approximations
$$
\left\langle \left|c_{i k}^{(m)}\right|^4\right\rangle\approx \frac{\langle r^2\rangle}{\langle r\rangle^2}f_{ik}^2
$$
and 
$$
\left\langle \left|c_{i k}^{(m)}\right|^2  \left|c_{i' k'}^{(m')}\right|^2\right\rangle\approx \frac{\eta}{\eta-1} \left(1-\frac{\langle r^2\rangle}{\langle r\rangle^2}\frac{1}{\eta}\right) f_{ik}f_{i'k'},
$$         
where we have used the shorthand notation $f_{ik}=f(E_{ik})$, $\langle r^n\rangle$ are the $n$-th moments of the probability distribution $p(r)$ and 
$$
\eta\equiv\frac{1}{\sum_{ikm} f_{ik}^2}\approx \frac{1}{\int \frac{\rho^2(E)}{\nu(E)}dE} 
$$
is the effective dimension of states  available for the ensemble.

With the previous approximations and following a similar procedure as \cite{Lerma-Hernandez(2019)} we obtain the following expression for the ensemble average of the asymptotic value of the survival probability
$$
\langle S_P^\infty\rangle_a=  \frac{\langle r^2\rangle}{\langle r\rangle^2}\frac{1}{\eta}+\frac{\left(1-\frac{\langle r^2\rangle}{\langle r\rangle^2}\frac{1}{\eta} \right)}{\eta-1}\sum_{i=1}^{N_e}d_i(d_i-1)\frac{\nu_i}{\nu},
$$  
whereas for the entire survival probability, we obtain the following expression for  the ensemble average
\begin{align}
&\langle S_P(t)\rangle_a=\\
&\frac{\left(1-\langle S_P^\infty\rangle_a \right)}{\eta-1}
\left[ \eta S_P^{bc}(t)-\sum_{i=1}^{N_e}d_i^2 \frac{\bar{\nu_i}}{\bar{\nu}}b_2\left(\frac{t}{2\pi \bar{\nu_i}}\right) \right]+
\langle S_P^\infty\rangle_a,\nonumber
\end{align}
where $S_P^{bc}(t)=\left|\int \rho(E) e^{-i E t}\right|^2$ gives the initial decay of the survival probability and $\bar{\nu_i}$ is the respective  mean  density of states in the energy window.   It is straightforward to show that the previous expression for the survival probability has the right value in $t=0$, $\langle S_P(t=0)\rangle_a=1$.  

For the  Bose-Hubbard model we consider in this paper ($L=N=9$), we have 6 sequences of energy levels, four of them have degeneracy two ($d_1=d_2=d_3=d_4=2$) and  the other two are non degenerate $d_{9-even}=d_{9-odd}=1$. The respective Density of States are $\nu_1=\nu_2=\nu_3=\nu_4=\nu/9$ and $\nu_{9-even}=\nu_{9-odd}=\nu/18$.
For this case, the  ensemble averages read
\begin{equation}
\langle S_P^\infty\rangle_a=\frac{\langle r^2\rangle}{\langle r\rangle^2}\frac{1}{\eta}+ \frac{\left(1-\frac{\langle r^2\rangle}{\langle r\rangle^2}\frac{1}{\eta} \right)}{\eta-1}\frac{8}{9}
\label{eq:AsSP-BH}
\end{equation}
and
\begin{align}
&\langle S_P(t)\rangle= \langle S_P^\infty\rangle_a +  \label{eq:AverSP-BH}\\
&\frac{\left(1-\langle S_P^\infty\rangle_a \right)}{\eta-1} 
\left[\eta S_P^{bc}(t)-\frac{16}{9}b_2\left(\frac{9 t }{2\pi \bar{\nu}}\right)-\frac{1}{9}b_2\left(\frac{9 t }{\pi \bar{\nu}}\right)\right]. \nonumber  
\end{align}
For the particular  uniform distribution $p(r)$ we consider in the main text 
   $\frac{\langle r^2\rangle}{\langle r\rangle^2}=\frac{4}{3}$, by substituting this ratio in Eqs.(\ref{eq:AsSP-BH}) and (\ref{eq:AverSP-BH}) we retrieve Eqs.(\ref{eq:span2}) and (\ref{eq:spasyman2}) of the main text.

\end{document}